\documentclass[11pt]{article}
\usepackage{natbib,amssymb,amsmath,graphicx,fullpage}
\usepackage[dvipsnames]{xcolor}
\usepackage{tikz}
\usepackage[hyphenbreaks]{breakurl}
\usepackage{color}
\PassOptionsToPackage{hyphens}{url}\usepackage[colorlinks,
            linkcolor=blue,
            urlcolor=magenta,
            citecolor=blue]{hyperref}

\newcommand{\bl}[1]{{\mathbf #1}}
\newcommand{\bs}[1]{{\boldsymbol #1}}

\newcommand{\mtr}[1]{\mathbf{#1}}
\newcommand{\vtr}[1]{\mathbf{#1}}
\newcommand{\mg}[1]{\boldsymbol{#1}}

\title{Multiplicative Coevolution Regression Models for Longitudinal Networks and Nodal Attributes}

\author{\begin{minipage}{0.5\textwidth}
\centering
Yanjun He\\
Department of Statistics\\
University of Washington\\
\end{minipage}
\begin{minipage}{0.5\textwidth}
\centering
Peter D. Hoff\\
Department of Statistical Science\\
Duke University\\
\end{minipage}}
\date{\today}

\begin{document}

\maketitle

\begin{abstract}
We introduce a simple and extendable coevolution model for the analysis of longitudinal network and nodal attribute data. The model features parameters that describe three phenomena: homophily, contagion and autocorrelation of the network and nodal attribute process. Homophily here describes how changes to the network may be associated with between-node similarities in terms of their nodal attributes. Contagion refers to how node-level attributes may change depending on the network. The model we present is based upon a pair of intertwined autoregressive processes. We obtain least-squares parameter estimates for continuous-valued fully-observed network and attribute data. We also provide methods for Bayesian inference in several other cases, including ordinal network and attribute data, and models involving latent nodal attributes. These model extensions are applied to an analysis of international relations data and to data from a study of teen delinquency and friendship networks.  

\end{abstract}

\section{Introduction}
Modern studies of social networks often involve longitudinal measurements
over time.
Such data can be represented as a sequence of sociomatrices
$\bl Y_0,\ldots, \bl Y_n$, where
each $\bl Y_t$ is a square $m\times m$ matrix with 
entry $y_{ij,t}$ representing
the value of a relationship between nodes $i$ and $j$ at time $t$
(the diagonal entries are typically undefined). 
Several methods for the analysis of such data have been developed:
Important early work in this area has involved stochastic
actor-oriented models \citep{snijders_2005,Snijders2010}.
This approach is based on
an economic model of rational choice, whereby individuals make unilateral
changes to their networks in order to maximize personal utility functions.
Other methods for dynamic network analysis have evolved out
of earlier methods for static network data. For example,
methods based on temporal
exponential random graph models (TERGM)
  have been developed
based on the popular static exponential random graph modeling framework
(ERGM) \citep{hunter_etal_2008,krivitsky_handcock_2014}.
An alternative approach to static network modeling 
is one where 
network patterns  are represented with node-specific latent variables
\citep{Nowicki2001,Hoff2002}.
Dynamic versions of these models have been
developed in
\citet{sarkar_moore_2005,xing_fu_song_2010,ward_ahlquist_rozenas_2013,Durante2014,Sewell2015}, among others.

Longitudinal network data will often be accompanied by
longitudinal node-level attributes $\bl X_0,\ldots, \bl X_n$,
where each $\bl X_t$ is an $m\times p$ matrix whose  $i$th row
is a vector $\bl x_{i,t}$ of characteristics of node $i$ at time $t$.
In such cases, it is often of interest to infer how
the network and nodal attributes might influence each other over time.
To this end, statistical methodology and software have
been developed that extends the actor-oriented approach described above
(\citet{snijders_steglich_schweinberger_2007}, \url{http://www.stats.ox.ac.uk/~snijders/siena/}).
While this work has been groundbreaking, the applicability of an
actor-oriented model
may be limited to certain types of networks and individual-level
characteristics.
As described by the primary developers of this approach
\citep{snijders_steglich_schweinberger_2007}, such a model may not be
appropriate in situations where network and behavioral data depend on
unobserved latent variables.
Such a situation may be present in
the study of social networks and obesity:
An individual's body mass index 
may be related to their social network, but
this relationship is likely mediated by other variables such as
socioeconomic status,  diet, exercise, participation in sports
and other variables that may potentially be unobserved.
Furthermore, parameter estimation for such actor oriented models
is computationally intensive, involving an iterative optimization
scheme that requires simulation of hypothetical networks at each
iteration.

As an alternative to this actor-oriented approach,
in this article we develop a class of coevolution models
for network and nodal attribute data that are based on
simple and scalable linear regression and latent factor models.
Like regression modeling,
the  framework we present is flexible and extendable, and
can be modified to accommodate  continuous and ordinal measurements for both
  the nodal  and network data.
The framework is built upon a simple autoregressive
model that describes the 
association of both  the
network $\bl Y_t$ and the nodal attributes $\bl X_t$ at time $t$
with the values $\{ \bl Y_{t-1}, \bl X_{t-1}\}$ from the previous
time point. The associations are modeled in terms of products 
of the network and nodal outcomes, and so we refer to such 
models as multiplicative coevolution regression (MCR) models. 

As we discuss in the next section, the parameters of MCR models 
can quantify three important data features:
First, that both the network and nodal attributes may vary smoothly
from time point to time point; second, the relations
between individuals may be influenced by the similarity of their attributes;
and third, individuals may change their attributes based upon the
attributes of those with whom they relate.
We refer to these three features as autocorrelation,
homophily, and contagion, respectively.
While the basic MCR model may simply be represented as a
type of regression model,
in Section 2 we discuss extensions of this model to accommodate 
network and nodal data that may be 
binary or ordinal, as well as extensions 
for
data where certain types of network patterns may be well-represented
with latent nodal factors.
In Section 3 we discuss estimation and inference, including 
maximum likelihood estimates for fully observed continuous data, and 
Bayesian inference for a variety of model extensions. 
In Section 4 we present two case studies. The first involves
monthly interactions between 50 countries over a 10 year period.
The second analyzes the coevolution of friendship ties
and an ordinal measure of delinquency for 26 high-school
students. A discussion follows in Section 5.

\section{Multiplicative Coevolution Regression}
A coevolution model for dynamic network and nodal attribute data should be able to quantify \emph{autocorrelation}, \emph{homophily} and \emph{contagion}.
Autocorrelation quantifies the tendency for
relations and attributes to vary gradually over time.
Homophily refers to the
possibility that changes to the relations between nodes may be partly determined by how similar their attributes are. Contagion describes how nodes may change their attributes based on the attributes of those with whom they have relations. 
For the case of undirected relational data, we propose the following simple multiplicative regression model for describing these three phenomena:
\begin{align}
y_{ij,t+1}&= \mu_{ij}  +\alpha y_{ij,t}+\bl x_{i,t}^T\bl H\bl x_{j,t}+\epsilon_{ij,t+1}, \label{eqn:mcrmodel} \\
\bl x_{i,t+1}&=\bs \theta_i + \bl A \bl x_{i,t}+\bl C\bl X_{t}^T\bl y_{i\cdot,t}+\bl e_{i,t+1}, \nonumber
\end{align}
where 
$\bl y_{i\cdot,t}$ is the $i$th row $\bl Y_t$ (with $y_{ii,t}=0$), 
the $\epsilon_{ij,t}$'s  are i.i.d.\ $N(0,\sigma^2)$ 
and the 
$\bl e_{i,t}$'s are  i.i.d.\ $N(\bl 0,\bs\Sigma)$. 
Alternatively, the intercept terms $\mu_{ij}$ and $\bs \theta_i$ 
can be replaced with regression terms 
involving exogenous predictors and 
possibly depending on time.

The parameters $\{\alpha,\bl A \}$, $\bl H$ and $\bl C$
respectively represent the phenomena of autocorrelation, homophily and
contagion described above.  To see this, note that  
if $\bl H$ and $\bl C$ were zero, 
then the model reduces to two first order autoregressive models, 
with $\alpha$ and $\bl A$ being the autoregression parameters. 
Regarding homophily, 
the matrix 
$\bl H\in \mathbb R^{p\times p}$ represents the influence of the similarity between the characteristics of two nodes on their relations. As a simple example, consider the case where $\bl H=h\bl I$ with $h>0$, and so $\bl x_{i,t}^T\bl H\bl x_{j,t}=h\bl x_{i,t}^T\bl x_{j,t}$. In this case we have positive homophily, in that the more similar $i$ and $j$ are in terms of their attributes at time $t$, the larger  the
expected relation between them at the next time point. 
Finally, the matrix $\bl C$ describes contagion, the 
effect of nodal attributes at time $t$ on those of a given node $i$ 
at time $t+1$, weighted by the relations of node $i$.  
For example, 
assume for the moment that  $y_{ij,t}\in \{0,1\}$. In this 
case, $\bl X_t^T\bl y_{i\cdot,t}$ is proportional to the 
average of the characteristic values of those to whom node $i$ is linked.

Model (\ref{eqn:mcrmodel}) describes the simplest situation we consider in this article, in which the network and nodal attributes are assumed to be Gaussian and fully observed.
We refer to this model as a multiplicative coevolution regression (MCR) model. 
The model is multiplicative in $\bl Y_t$ and $\bl X_t$ via the 
homophily and contagion effects. However, as will be discussed in 
Section 3, it is linear in the parameters and so 
can be viewed as a multivariate linear regression model.   

The assumption of additive effects and normally distributed outcomes
is not appropriate for many network datasets. 
In particular, many network relations are binary or ordinal,
and are possibly asymmetric in that $y_{ij,t}$ is not necessarily equal
to $y_{ji,t}$. Furthermore, it is often likely to be the case that
some variables that drive network formation are unobserved, and
not part of the the dataset.  In this case, we may want to
augment the model  to accommodate
latent, unobserved nodal characteristics. We consider extensions
of the model in (\ref{eqn:mcrmodel}) to accommodate each of these
situations in the  following paragraphs.

\paragraph{Ordinal data:}
The relational variable $y_{ij,t}$ in many 
network datasets is binary, indicating whether or 
not two nodes have some sort of tie between them, 
such as friendship  or
social interaction. In other cases this  variable is ordinal, such as
when $y_{ij,t}$ is  recorded as  being
negative, neutral or positive,
or when $y_{ij,t}$   measures the number or intensity
of social interactions between
two individuals.
While the assumptions of Gaussian noise and additive effects of the MCR model
will not generally be appropriate
for such data, the model can be used to formulate a
probit regression model for general ordinal network relations.
This is done by expressing  the 
relations $y_{ij,t}$ 
as a non-decreasing function of latent relations $z_{ij,t}$ that 
that do follow  the MCR model. Specifically, we assume that 
$y_{ij,t}= f(z_{ij,t})$ for some non-decreasing function $f$, 
and that the process $\{ (\bl Z_t,\bl X_t):t=0,\ldots,n\}$ follows
the Gaussian MCR model. The only adjustment to the model is 
that the error variance $\sigma^2$ may be assumed to be 1, as 
otherwise this scale parameter is not separately identifiable from
$f$. 
Furthermore, if the nodal characteristic process $\{ \bl X_t ,t=1,\ldots,n\}$ is not
well represented with a normal model then an ordinal 
probit model can be used here as
well. 
In this case, we model $x_{i,k,t} = g_k(w_{i,k,t})$ where 
$g_1,\ldots, g_p$ are nondecreasing functions and 
$w_{i,k,t}$ is a latent Gaussian process that determines 
$x_{i,k,t}$. 
Letting $\bl W_t$ be  the $n\times p$ matrix with elements $w_{i,k,t}$, 
the model is completed by assuming   
$\{ (\bl Z_t,\bl W_t):t=0,\ldots,n\}$ follows the MCR model. 
An example data analysis in which the both the relational and 
attribute variables are ordinal is presented in Section 4.

\paragraph{Directed relations:}
Many network datasets include directed relations where $y_{ij,t}$ is not necessarily equal to $y_{ji,t}$. The natural extension of the multiplicative coevolution model in equation (\ref{eqn:mcrmodel})
to accommodate directed relations is as follows:
\begin{align}
y_{ij,t+1}&= \mu_{ij}  +\alpha_1 y_{ij,t}+\alpha_2 y_{ji,t} + \bl x_{i,t}^T\bl H\bl x_{j,t}+\epsilon_{ij,t+1}, \label{eqn:obsmodel} \\
\bl x_{i,t+1}&=\bs \theta_i + \bl A \bl x_{i,t}+\bl C_1\bl X_{t}^T\bl y_{i\cdot,t}+ \bl C_2\bl X_{t}^T\bl y_{\cdot i ,t} +   \bl e_{i,t+1}. \nonumber
\end{align}
The modifications to the model for the network  process are that
the homophily parameter $\bl H$ is not necessarily symmetric, and that $y_{ij,t+1}$   may be influenced
by $y_{ji,t}$ via the reciprocity parameter $\alpha_2$. 
The model for the attribute process now includes
two different contagion parameters $\bl{C}_1$ and $\bl{C}_2$. The former represents the relationship-weighted effect of the nodal characteristics of those to which one sends ties, while the latter represents the effect of those from which one receives ties. An example data analysis using a probit version of this directed 
MCR model appears in Section 4.

\paragraph{Latent nodal attributes:}
When nodal attribute data 
are either not available or only weakly associated with the network process, it may be useful to  add latent nodal attributes to the model.
In the case of static network modeling, 
 inclusion of latent nodal attributes can provide
identification of clusters of nodes, 
improved model fit and better
out-of-sample predictions of unmeasured relations. 
The basic framework is to model the relation
$y_{ij}$ between nodes $i$ and $j$ as depending
on the similarity of latent, unobserved characteristics $\bl x_i$ and 
$\bl x_j$. For example, 
the latent class model of \citet{Nowicki2001}  is equivalent to 
letting $\bl x_i$ represent a vector indicating membership of node 
$i$ to one of several latent classes.  The latent distance model 
of \citet{Hoff2002} assumes $y_{ij}$  depends on the Euclidean 
distance between the latent location vectors $\bl x_{i}$ and $\bl x_j$. 
\citet{Hoff2008} shows how both of these approaches are 
generalized by a multiplicative approach, in which 
 $y_{ij}$  is modeled 
as a function of the inner product $\bl x_i^T \bl H \bl x_j$.  
This suggests that, in the absence of nodal characteristics strongly associated 
with the network process, we allow $\bl x_{i,t}$ in 
the MCR model (\ref{eqn:mcrmodel})  to represent latent, unobserved nodal 
attributes. 
In this case, both the parameters of the MCR model in 
(\ref{eqn:mcrmodel}) and 
the latent attribute process $\{ \bl X_t: t=0,\ldots,n\}$
can be estimated from the data. 
However, the parameters in the MCR model 
are not fully identifiable when the nodal attributes are latent.  
For example, the model is invariant to orthogonal rotations of 
the $\bl X_t$'s, that is, replacement of each $\bl X_t$ by 
$\bl X_t \bl R$, where $\bl R$ is a $p\times p$ orthogonal  
matrix so that $\bl R\bl R^T=\bl I$.  
For this reason we simplify the latent MCR model by 
parameterizing the homophily parameter $\bl H$ as being diagonal, 
and setting $\bs\Sigma$ equal to the $p\times p$ identity matrix. 
Even so, the model remains invariant to simultaneous permutations of the columns 
of the $\bl X_t$'s. This issue is discussed further in the data analysis 
example in Section 4. 

\begin{figure}[!ht]
\centering
\begin{tikzpicture} [
  auto,
  block/.style    = { shape=circle, draw=black, thick,
                      text width=1.85em, text centered},
  ]
\matrix [column sep=7mm, row sep=5mm] {
 \node (null1) {}; & \node [block] (Y1) {$\bl{Y}_{t-1}$} ; &\node [block] (Y2) {$\bl{Y}_t$}; & \node [block] (Y3) {$\bl{Y}_{t+1}$}; &\node (null2) {}; \\
 \node (null3) {}; & \node [block] (X1) {$\bl{X}_{t-1}$} ; &\node [block] (X2) {$\bl{X}_t$}; & \node [block] (X3) {$\bl{X}_{t+1}$}; &\node (null4) {}; \\
  };
\path [thick, ->] (null3) edge (X1);
\path [thick, ->] (X1) edge (X2);
\path [thick, ->] (X2) edge  (X3);
\path [thick, ->] (X3) edge  (null4);
\path [thick, ->] (X1) edge  (Y1);
\path [thick, ->] (X2) edge  (Y2);
\path [thick, ->] (X3) edge  (Y3);

\end{tikzpicture}
\begin{tikzpicture} [
  auto,
  block/.style    = { shape=circle, draw=black, thick,
                      text width=1.85em, text centered},
  ]
\matrix [column sep=7mm, row sep=5mm] {
 \node (null1) {}; & \node [block] (Y1) {$\bl{Y}_{t-1}$} ; &\node [block] (Y2) {$\bl{Y}_t$}; & \node [block] (Y3) {$\bl{Y}_{t+1}$}; &\node (null2) {}; \\
 \node (null3) {}; & \node [block] (X1) {$\bl{X}_{t-1}$} ; &\node [block] (X2) {$\bl{X}_t$}; & \node [block] (X3) {$\bl{X}_{t+1}$}; &\node (null4) {}; \\
  };
\path [thick, ->] (null1) edge (Y1);
\path [thick, ->] (Y1) edge (Y2);
\path [thick, ->] (Y2) edge  (Y3);
\path [thick, ->] (Y3) edge  (null2);
\path [thick, ->] (null3) edge (X1);
\path [thick, ->] (X1) edge (X2);
\path [thick, ->] (X2) edge  (X3);
\path [thick, ->] (X3) edge  (null4);
\path [thick, ->] (X1) edge  (Y2);
\path [thick, ->] (X2) edge  (Y3);
\path [thick, ->] (Y1) edge  (X2);
\path [thick, ->] (Y2) edge  (X3);

\end{tikzpicture}
\caption{Dependence graphs for longitudinal network models. Hidden Markov model (left) and latent MCR model (right). }
\label{figure:diagram}
\end{figure}
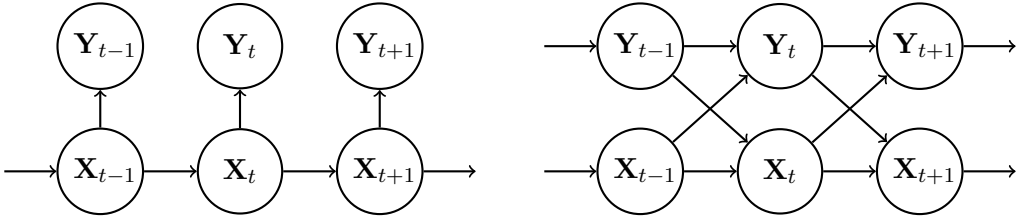

This latent MCR model is similar to several other models developed 
for the analysis of longitudinal network data that lack nodal 
attributes. For example, \citet{ward_ahlquist_rozenas_2013, Durante2014} 
and \citet{Sewell2015} 
each utilize models where the network $\bl Y_t$ at each time point 
is modeled as a function of nodal latent variables $\bl X_t$, which 
in turn follows a stochastic process. These  
are hidden Markov models for the observed network process, 
and can be graphically depicted by the dependence graph in the first panel of 
Figure \ref{figure:diagram}.  
Such models essentially only include a homophily parameter, modeling 
a relation between two nodes as a function of their time-varying 
latent attributes. 
In contrast, our latent MCR model 
(depicted in the second panel of the figure) permits 
a richer description of the evolution 
of the network by inclusion of an autocorrelation term for the 
network, and a contagion parameter that allows for the possibility that 
nodes may change their nodal attributes depending on their 
past relations.

\section{Estimation and Inference} 
One feature of the MCR model is its simplicity: It can be expressed as a
pair of linear regression models. 
As we show in the next subsection, this permits very easy 
parameter estimation  in the case of a normal model 
for the observed network and attributes. 
The linear regression 
framework also 
serves as a building block for
data analysis in more complicated 
situations, such as the case of  ordinal relational and attribute variables
and latent attribute models. As we show in Section 3.2, 
Bayesian inference in such situations can be obtained using 
relatively straightforward Gibbs sampling algorithms.

\subsection{MLEs for normal models}
To see how the network evolution model in Equation  \ref{eqn:mcrmodel}
can be expressed as a linear regression model,  
first parameterize 
$\mu_{ij}$ as $\mu_{ij} = \bs\gamma^T \bl s_{ij}$, where 
$\bl s_{ij}$ is  a vector of observed exogenous covariates 
and $\bs \gamma$ is a vector of unknown parameters. 
If there are no exogenous covariates then  we can take  
$\bs \gamma$ to simply be a vector consisting of the values of
$\mu_{ij}$ and
$\bl s_{ij}$ to be the appropriate binary vector with a single entry
equal to one and the remaining entries equal to zero.
Then, note that the term 
$\bl x_{i,t}^T \bl H \bl x_{j,t}$ can be written  
as $\bl h^T \bl x_{ij,t}$, where 
$\bl h = \text{vech}(\bl H)$ is the ``half vectorization''
of the matrix $\bl H$ obtained by concatenating the lower-triangular 
elements of $\bl H$ (including the diagonal), 
and $\bl x_{ij,t} = 
   \text{vech}( \bl x_{i,t} \bl x_{j,t}^T +
                \bl x_{j,t} \bl x_{i,t}^T - \text{diag}( \bl x_{i,t} \bl x_{j,t}^T ))$. For example, if each $\bl x_{i,t}$ is two-dimensional, 
then  $\bl x_{ij,t} = ( x_{i,1,t} x_{j,1,t},  \,
                          x_{i,1,t} x_{j,2,t} + x_{i,2,t} x_{j,1,t}, \,
   x_{i,2,t} x_{j,2,t} )$. 
We can therefore write the network component of  model (\ref{eqn:mcrmodel})
as 
\[
 y_{ij,t} = \bs \beta^T\bl w_{ij,t} + \epsilon_{ij,t} , 
\]
where $\bs\beta = ( \bs\gamma, \alpha, \bl h)$ and 
$\bl w_{ij,t} 
= ( \bl s_{ij} , y_{ij,t-1} , \bl x_{ij,t} )$.  
The residual sum of squares can be expressed as 
\[
\sum_{t} \sum_{i<j} ( y_{ij,t} - \bs\beta^T\bl w_{ij,t})^2 = 
\left (  \sum_{t} \sum_{i<j}  y_{ij,t}^2 \right )  - 2\bs \beta^T\bl l + 
  \bs \beta^T \bl Q \bs \beta, 
\]
where
\begin{align}
\bl l & = \sum_{t=1}^n \sum_{i<j} \bl w_{ij,t}^T  y_{ij,t}  
\label{eqn:netlq} \\
\bl Q & = \sum_{t=1}^n \sum_{i<j} \bl w_{ij,t} \bl w_{ij,t}^T.  \nonumber
\end{align}
The maximum likelihood  estimate of $\bs \beta$ is therefore given by 
$\hat{\bs\beta} = \bl Q^{-1} \bl l$.  

The attribute evolution model is also a linear regression model. 
Parameterizing $\bs\theta_i$ as $\bs \Gamma \bl s_i$ for an exogenous 
covariate vector $\bl s_i$ and parameter matrix $\bs\Gamma$, we have 
$\bl x_{i,t+1} = \bl B \bl w_{i,t+1} + \bl e_{i,t+1}$, 
where $\bl B$ is the column-wise concatenation of 
$\bs\Gamma$, $\bl A$ and $\bl C$, and 
$\bl w_{i,t+1}$ is the vector obtained by concatenating  the vectors
$\bl s_i$, $\bl x_{i,t}$ and $\bl X_t^T \bl y_{i\cdot, t}$.   
The attribute evolution model can be written in matrix form 
as 
\begin{align*}
  \bl X_{t+1} &= \bl S \bs \Gamma^T + \bl X_t \bl A^T + \bl Y_{t} \bl X_{t} 
  \bl C^T  + \bl E_{t+1}  \\ 
      &=  \bl W_{t+1} \bl B^T  + \bl E_{t+1}, 
\end{align*}  
where the $i$th row of $\bl W_{t+1}$ is the vector $\bl w_{i,t+1}$
defined above. 
A standard result from multivariate regression is that the
MLE of $\bl B$ is given by $\hat {\bl B} =\bl L \bl Q^{-1}$, where 
\begin{align}
\bl L & = \sum_{t=1}^n  \sum_{i=1}^m \bl x_{i,t} \bl w_{i,t}^T 
    =   \sum_{t=1}^n \bl X_{t}^T \bl W_{t} \label{eqn:attlq}  \\
\bl Q  & =  \sum_{t=1}^n  \sum_{i=1}^m \bl w_{i,t} \bl w_{i,t}^T  
      =   \sum_{t=1}^n \bl W_t^T \bl W_t .  \nonumber
\end{align}

Estimation for directed relations proceeds with a few modifications. 
For estimation of the network process, 
$\bs \beta = (\bs \gamma, \alpha_1,\alpha_2, \bl h)$
where $\bl h= \text{vec}(\bl H)$, and 
$\bl w_{ij,t} 
= ( \bl s_{ij} , y_{ij,t-1} , y_{ji,t-1}, \bl x_{j,t}\otimes \bl x_{i,t})$, 
where ``$\otimes$'' is the Kronecker product. 
Additionally, the 
summation in (\ref{eqn:netlq})  is replaced by a summation over 
all ordered pairs $\{ (i,j): i\neq j\}$. 
For estimation of the attribute process, 
the matrix $\bl B$ is the concatenation of $\bs \Gamma$, $\bl A$, $\bl C_1$
and $\bl C_2$, and $\bl w_{i,t+1}$ is the concatenation  of  the vectors
$\bl s_i$, $\bl x_{i,t}$, $\bl X_t^T \bl y_{i\cdot, t}$ and 
  $\bl X_t^T \bl y_{\cdot i, t}$.

\subsection{Bayesian estimation for model extensions}  
In cases where nodal attributes are not observed or 
the network and attribute processes are not plausibly 
Gaussian, the MCR model will have to be extended as 
described in Section 2. For these cases we propose 
a Bayesian approach to inference, as a posterior 
approximation scheme based on Gibbs sampling is 
modular and can be easily modified to accommodate 
different features of the data.  We first discuss Bayesian 
inference for the basic MCR model described in 
Equation \ref{eqn:mcrmodel}, and then discuss 
two modifications, permitting the modeling of 
unobserved latent attributes and the modeling 
of ordinal network and attribute data. 

Let
$\bs \beta= (\gamma, \alpha, \bl h)$ and
$\bl B = [ \bs \Gamma \, \bl A \, \bl C ] $ be the regression parameters
in the network and attribute processes respectively. 
Using semiconjugate  prior distributions for the unknown parameters
$\bs \beta$,  $\bl B$, $\sigma^2$ and $\bs\Sigma$,  
their joint posterior distribution can be 
approximated with a Gibbs sampler that iteratively simulates
values of these parameters from their full conditional distributions.
Specifically, if the prior distributions are
$\bs \beta \sim N( \bl 0 , \bl V_{\beta})$,
$\bl b = \text{vec}( \bl B)  \sim N ( \bl 0 , \bl V_{\rm b})$, 
$1/\sigma^2 \sim $ gamma$(\nu_0/2,\nu_0\sigma_0^2/2)$, 
and $\bs\Sigma^{-1} \sim $Wishart$(\bl S_0^{-1} , \eta_0)$, 
then the
Gibbs sampler proceeds by iterating the following steps:
\begin{enumerate}
\item Simulate $\bs \beta$ from its multivariate
normal full conditional distribution
with mean $( \bl V_{\beta}^{-1} + \bl Q )^{-1} \bl l$ and
variance  $( \bl V_{\beta}^{-1} + \bl Q )^{-1}$, where
$\bl Q$ and $\bl l$ are as in (\ref{eqn:netlq}). 
\item Simulate $\bl b$ 
from its multivariate
normal full conditional distribution
with mean $( \bl V_{b}^{-1} +  \bl Q \otimes  \bs\Sigma^{-1} )^{-1} \text{vec}(\bl L)$ and
variance  $( \bl V_{b}^{-1} + \bl Q \otimes \bs\Sigma^{-1} )^{-1}$, where
$\bl Q$ and $\bl L$ are as in (\ref{eqn:attlq}).

\item  Simulate $1/\sigma^2 \sim $ gamma$( [\nu_0 + n m(m-1)/2]/2, 
          [\nu_0 \sigma^2_0 + RSS  )  ]/2)   $, where
\[RSS = \sum_{t=1}^n \sum_{i<j} (y_{ij,t} - 
 [ \mu_{ij} + \alpha y_{ij,t-1} + \bl x_{i,t-1}^T \bl H \bl x_{j,t-1} ] )^2.  
 \] 
\item Simulate $\bs \Sigma^{-1} \sim $ Wishart$( [  \bl S_0 + {\bf RSS} ]^{-1} , 
    \eta_0 + mn )$, where   
\[
{ \bf RSS} = \sum_{t=1}^n ( \bl X_t -  
     [ \Theta + \bl X_{t-1} \bl A^T +  \bl Y_{t-1} \bl X_{t-1}\bl C^T] )^T 
(\bl X_t -  [ \Theta + \bl X_{t-1} \bl A^T +  \bl Y_{t-1} \bl X_{t-1}\bl C^T]) . 
\]
\end{enumerate}
Iteration of this algorithm generates a Markov chain with a 
stationary distribution equal to the posterior 
distribution of $(\bs\beta, \bl b, \sigma^2 , \bs \Sigma)$. 
The empirical distribution of the simulated parameter values 
can be used to obtain approximate posterior means, quantiles 
and confidence intervals. 
Furthermore, the 
Gibbs sampling algorithm can be modified or extended to 
provide inference for related models and data structures. 
We consider two such modifications below. 

\paragraph{Latent attribute models:}
The Gibbs sampling algorithm may be easily modified to 
accommodate the case 
that the $\bl x_{i,t}$'s are estimated latent attributes
rather than observed attributes. 
Recall from the discussion in Section 2 that in this 
case we fix $\bs \Sigma = \bl I$ for reasons of identifiability. 
As such, we replace Step 4 in the Gibbs sampler described above
with the 
following step 
that iteratively simulates values of the $\bl x_{i,t}$'s  
from their full conditional distributions:
\begin{enumerate}
\item[4.] Iteratively over nodes $i=1,\ldots,m$ and time points
$t=0,\ldots,n$, simulate $\bl x_{i,t}$ from its multivariate normal full
conditional distribution. For a time point $t$ such that $0<t<n$, this
full conditional distribution has mean $\bl Q^{-1} \bl l$ and variance
$\bl Q^{-1}$, where
$\bl l = \sum_{k=1}^3 \bl W_k^T \bl z_k $ and
$\bl Q = \sum_{k=1}^3 \bl W_k^T \bl W_k$ are given as follows:
\begin{align*} 
\bl W_1  = \bl I & & 
\bl z_1 = \bs \theta_i  + \bl A \bl x_{i,t-1} + 
    \bl C \bl X_{t-1}^T \bl y_{i\cdot,t-1}   \\
\bl W_2 =  \tilde{\bl X}_{t} \bl H/\sigma    & &  
\bl z_2 =   (  \tilde {\bl y}_{i\cdot,t+1} - 
     \tilde {\bs\mu}_{i\cdot}  - \alpha \tilde {\bl y}_{i\cdot, t} )/\sigma        \\
\bl W_3 =  \bl e_i\otimes \bl A +  {\bl y_{i\cdot} }_t \otimes \bl C  & &  
\bl z_3 = \text{vec}( \bl X_{t+1} - \Theta  -\tilde{\bl I}^T \tilde {\bl X}_t \bl A^T - \tilde {\bl Y}_t^T \tilde {\bl X}_t \bl C^T  ), 
\end{align*}
\end{enumerate} 
where 
$\bl e_i$ is a vector of zeros except for a one in the $i$th entry, and 
the tildes in the formulas for $\bl W_2,\bl W_3$ and $\bl z_2, \bl z_3$ indicate 
the removal of the $i$th row of a matrix or the $i$th element of a vector.   
The three terms in the sums for $\bl l$ and $\bl Q$ represent information 
about $\bl x_{i,t}$ 
from the past, from the future network, and 
from the future attributes, respectively. 
The values of $\bl x_{i,0}$ and $\bl x_{i,n}$ are updated similarly, 
except in the former case we have 
$\bl z_1 = \bl 0$, and in the latter case we have 
$\bl l =  \bl W_1^T \bl z_1 $ and
$\bl Q =  \bl W_1^T \bl W_1$. 
As discussed in Section 2, we also 
restrict $\bl H$ to be a diagonal matrix when the attributes are latent. 
As a result, the calculation of $\bl l$ in Step 1 of the Gibbs sampler
is 
as in (\ref{eqn:netlq})
except that it is computed with
$\bl x_{ij,t} =  (\bl x_{j,t}\circ \bl x_{j,t})$, where
``$\circ$'' denotes element-wise multiplication. This is because
$\bl x_{i,t}^T \bl H \bl x_{j,t}= (\bl x_{j,t}\circ \bl x_{j,t})^T \bl h$
in this case where $\bl H$ is diagonal.
A numerical illustration of this Gibbs sampler as applied to longitudinal 
international relations data is provided in Section 4.1.

\paragraph{Probit models for ordinal outcomes:}
Ordinal network and  attribute  data may be accommodated 
by modeling the observed network and attribute processes 
as non-decreasing functions of latent processes that 
do follow the Gaussian MCR model in Equation \ref{eqn:mcrmodel}. 
Specifically, 
let $y_{ij,t}$ be the observed ordinal-valued relation between 
nodes $i$ and $j$ at time $t$, and let 
$x_{i,k,t}$ be the value of the $k$th ordinal-valued attribute 
of 
node $i$ at time $t$. We then model the network and attribute 
process by assuming 
 $y_{ij,t} = f( z_{ij,t}) $ and 
$x_{i,k,t} = g_k(w_{i,k,t})$, where 
 $f$ and $g_1,\ldots, g_p$ are unknown non-decreasing step functions, 
and the $z_{ij,t}$'s and $w_{i,k,t}$'s follow the 
Gaussian MCR model. 
A Gibbs sampler for this probit MCR model may be obtained 
by adding to Steps 1-4 above a few additional steps to 
simulate values of the $z_{ij,t}$'s, the $w_{i,k,t}$'s 
from their full conditional distributions, as well as the 
values defining $f$ and $g_1,\ldots, g_p$. 
Such steps are standard in the literature on Bayesian modeling 
of ordinal data: 
Assuming normal prior distributions for the 
locations of the jumps in $f$, $g_1,\ldots, g_p$, 
the full conditional distributions of all of these 
quantities are constrained normal distributions, which may be 
simulated from using the inverse-CDF method. For information 
on such procedures in general, see 
\citet{albert_chib_1993}. Details of the Gibbs sampler 
for the MCR model in particular can be found in Appendix \ref{appendixA}. An example data 
analysis using the probit MCR model appears in Section 4.2.

\section{Example Data Analyses} 
In this section we illustrate the use of the 
MCR model with two example data analyses.  
The first example applies the model to 
a time series of international relations between 
50 countries over a ten year period using
a latent Gaussian MCR model.  
The second example studies the coevolution 
of the friendships and delinquency behaviors of 26 high-school students.  
In this latter example the network is binary and the nodal attribute 
is ordinal, and so an ordinal MCR model is employed. 

\subsection{International Relations} 
The ICEWS project 
(\url{http://www.lockheedmartin.com/us/products/W-ICEWS/iData.html})
gathers data on 
international events occurring between countries. 
For this article, we analyze a monthly summary of the 
undirected dyadic relations between the 50 most active countries in 
the ICEWS database during a 112 month period from  2006 to 2015.
Events between countries are assigned event codes, and each 
event has an associated intensity score ranging 
from -10 for extreme negative relations to +10 
for extreme positive relations \citep{ICEWS}. 
For this analysis, we computed the monthly sum of these intensity 
scores for each pair of countries, and then applied a normal quantile-quantile 
transformation to all values.  This resulted in a time series of 112 
$50\times 50$ sociomatrices $\bl Y_{0}, \ldots, \bl Y_{111}$, 
where $y_{ij,t}$ is the (transformed) intensity score sum 
between countries $i$ and $j$ for month $t$. 

We fit the  latent MCR model described in Section 2 
with $p=2$  latent 
attributes for each country at each time point. 
With all regression 
coefficients being a priori i.i.d.\ $N(0,100)$, and 
$\nu_0=\sigma^2_0=1$, the Gibbs sampler described in Section 3.2 
was run for 27,500 iterations.  The first 2,500 iterations of the 
algorithm were dropped to allow 
for burn-in, and every 10th iteration thereafter was saved, 
yielding 2,500 simulated values for 
each parameter with which to approximate the posterior distribution. 
The average effective sample size across parameters in the MCR model 
was 789.

A 95\% posterior credible interval for 
$\alpha$ is (0.134,0.145), indicating strong evidence for 
positive autocorrelation, and the diagonal values of $\bl H$ were 
positive for every iteration of the Gibbs sampler, indicating 
positive homophily. 
To get a sense of the magnitude of these coefficients, we computed the relative 
sum of squares contributions of the
four terms of the network coevolution model, averaged across 
time points. These contributions were 
28.2,  2.3,  16.2 and  53.2 percent, respectively, for
the $\mu_{i,j}$'s, the autoregressive term, the homophily term and 
the error variance, respectively. 

\begin{table}
\begin{center}
\begin{tabular}{r||cccc|cccc}  
 quantile& $a_{1,1}$ & $a_{1,2}$ & $a_{2,1}$  &  $a_{2,2}$ & 
           $c_{1,1}$ & $c_{1,2}$ & $c_{2,1}$  &  $c_{2,2}$   \\  \hline
2.5\% & 0.148 &0.044 &-0.004 &0.339 &0.005 &-0.002 &0.001& 0.015 \\
50\%  & 0.193 &0.094 & 0.047 &0.388 &0.006 & 0.000 &0.004& 0.018 \\
97.5\%& 0.241 &0.144 & 0.098 &0.438 &0.008 & 0.002 &0.006& 0.020
\end{tabular}
\end{center} 
\caption{Posterior quantiles of parameters in the attribute evolution process
for the ICEWS data.}
\label{tab:iceatt} 
\end{table}

Posterior medians and 95\% credible intervals for the $\bl A$ and $\bl C$
parameters of the attribute evolution model
are given in Table \ref{tab:iceatt}. The most significant terms 
in these two matrices are the diagonal terms, indicating that 
the two latent attribute processes both show  positive autocorrelation 
and positive contagion, but not strong interdependence with each other. 
The magnitude of the autocorrelation and contagion effects 
can be assessed by computing the sum of squares of these terms
 relative to the  $\bs \theta_i$'s and the error term, averaged across 
time points.  These contributions were
60.0,  9.3,  4.6 and  26.1 percent, respectively, for
the $\bs\theta_i$'s, the autoregressive term, the contagion term and
the error variance, respectively.

\begin{figure}
\centerline{\includegraphics[width=6.5in]{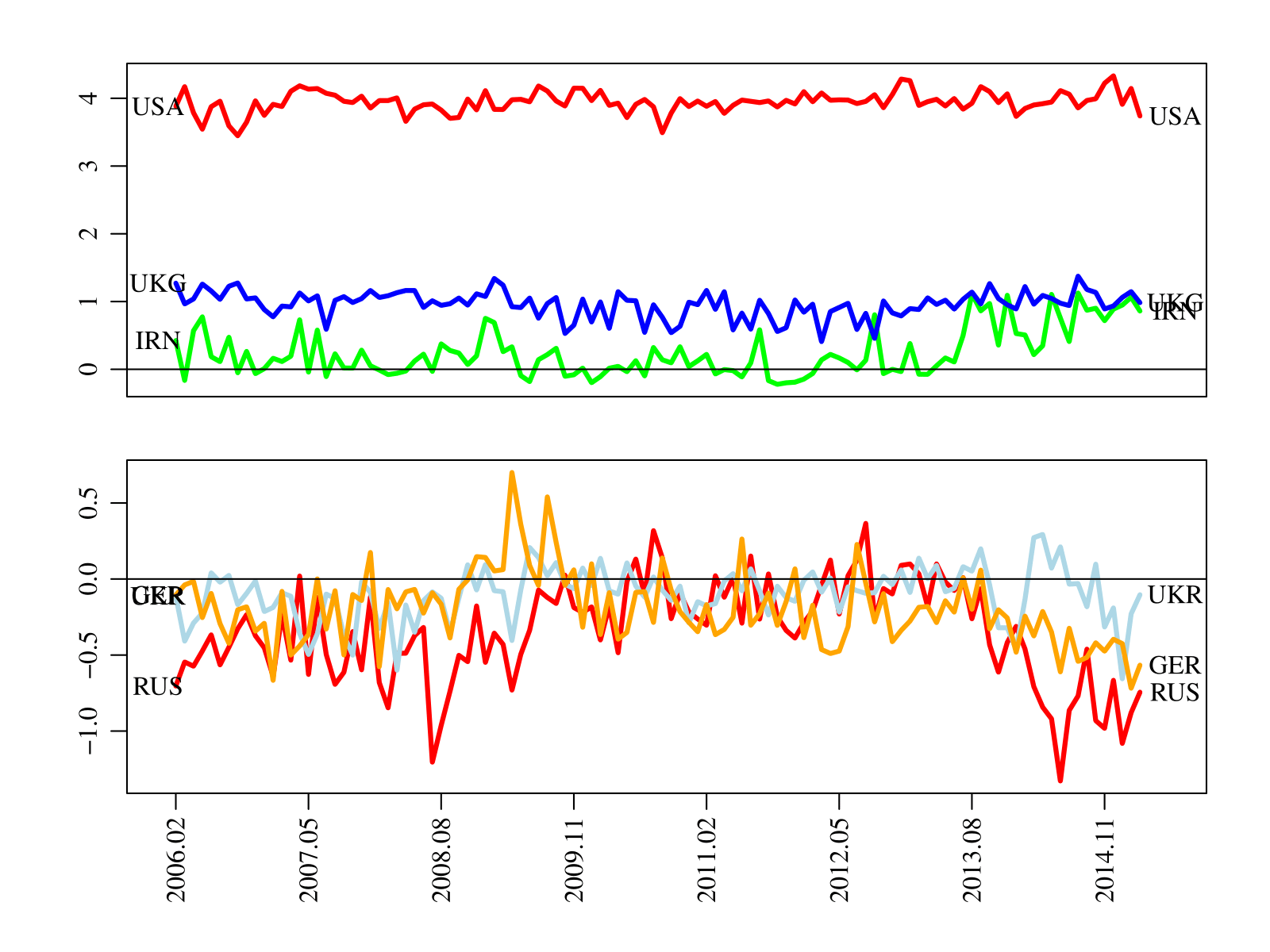}} 
\caption{Time series of selected country-specific latent attributes. 
The top plot gives the estimated values of the first factor 
for the United States (USA), Iran (IRN) and the United Kingdom (UKG). 
The lower plot
gives values of the second factor for Ukraine (UKR), Germany (GER) and 
Russia (RUS). } 
\label{fig:icews}
\end{figure}

Figure \ref{fig:icews} plots the times series of the estimated 
latent attributes for a few selected countries. The top panel 
plots the first attribute (corresponding to the larger 
of the two homophily effects) for the United States, the United Kingdom and 
Iran. The plot indicates that this factor contributes positively 
to the relationship between the United States and the United Kingdom 
throughout the time period (as the estimated attributes have the 
same sign), whereas the contribution to the relationships of these 
countries with Iran is neutral until early 2013, when 
Hassan Rouhani was 
elected President over several hardline candidates 
and indicated a desire to negotiate a nuclear accord. 
The second panel of the figure plots a time series 
of the second latent attribute for Ukraine, Germany and Russia. 
In this plot we see that the time series for Russia and Ukraine
are similar until the very beginning of 2014, when 
the protests against the Russian-backed government of 
President Yanukovych began.

Finally, we performed a small out-of-sample 
forecasting study to assess the benefit of 
the proposed model over the type of hidden Markov 
model considered in \citet{ward_ahlquist_rozenas_2013, Durante2014}, 
and displayed graphically in the left-hand side of Figure 
\ref{figure:diagram}. 
Such models lack the network autocorrelation term 
$\alpha$ and the contagion term $\bl C$. 
To assess the predictive benefit of these effects
we considered four models -  with and without $\alpha$
and with and without $\bl C$. We obtained five 
one-month-ahead forecasts  for each model, 
using data up to and including months 
87, 92, 97, 102, 107 to predict the value of 
the network at time 88, 93, 98, 103 and 108 respectively. 
In terms of prediction error sum of squares, 
the full MCR model with network autocorrelation 
and contagion effects performed the best for each month 
forecasted. However, the submodel without 
contagion effects only performed 1.5\% worse, on 
average over the five months forecasted. 
However, the submodel 
lacking both network autocorrelation and contagion 
performed on average 6.8\% worse, suggesting that for these 
data, network autocorrelation effects are more important 
than contagion effects for forecasting the network.

\subsection{Friendship and Delinquency} 
\citet{Knecht2007} gathered gathered data on a small directed friendship 
network of 25 Dutch secondary school students, along with 
nodal attributes including sex and a five-level ordinal 
measure of delinquency.  Both delinquency and the friendship network
were measured at four time points during a year-long period. 

We model the coevolution of friendship and delinquency  
over the study period with an ordinal MCR model. 
Specifically, 
we model the binary friendship indicator
$y_{ij,t}$  as $y_{ij,t} = f( z_{ij,t})$,  
and the delinquency category $x_{i,t}$  as
 $x_{i,t} = g(w_{i,t})$, where
$f$ and $g$ are non-decreasing functions and 
 $z_{ij,t}$ and  $w_{i,t}$ follow a Gaussian MCR model:
\begin{align*}
z_{ij,t+1}&= \bs \beta^T\bl s_{i,j}  +\alpha_1 z_{ij,t}+ \alpha_2 z_{ji,t} + h  w_{i,t}w_{j,t}+\epsilon_{ij,t+1}  \\ 
w_{i,t+1}&= b s_ i +  a  w_{i,t}+ c_1\bl w_{t}^T\bl z_{i\cdot,t}+ 
                         c_2\bl w_{t}^T\bl z_{\cdot i,t}+ e_{i,t+1} \\
 \{\epsilon_{i,j,t} \} , \{e_{i,t} \} & \sim \text{ i.i.d. }   N(0,1) , 
\end{align*}
where $\alpha_1$, $\alpha_2$ and $a$
describe network autocorrelation, network reciprocity, and 
delinquency autocorrelation respectively, 
$h$ is a homophily parameter and 
$c_1$ and $c_2$ are  contagion parameters. 
Additionally, $s_i$ is the binary indicator that student $i$ is female, 
 and  $\bl s_{i,j} = (s_i, s_j , 1(s_i=s_j))$ is a vector 
describing the gender characteristics of the directed dyad $(i,j)$.   
The unknown parameters $\bs \beta$ and $b$ 
describe the effects of gender on temporal changes to the  
network and the nodal 
attributes, respectively.  
We also note that the latent variables $z_{i,j,1}$ and $w_{i,1}$
at the first time point  were modeled  
as $z_{i,j,1} \sim N(\bs \gamma^T \bl s_{i,j}, \sigma^2) $ 
and $w_{i,1} \sim N( g  s_i , \tau^2)$, respectively. 
The parameters $\bs \beta_1$ and $b_1$ describe the effects 
of gender on the initial state of the network and delinquency. 

The parameters in this model were estimated using the 
Gibbs sampler for ordinal data 
described in Section 3. We ran the MCMC algorithm for 
40,000 iterations, and dropped the first 20,000 iterations to allow 
for burn-in of the Markov chain.  
The lowest effective sample size 
 among the regression parameters was 643, and the median 
effective sample size was around 3000. 
Posterior medians and 95\% posterior  credible 
intervals are given in Table \ref{table:case2coef}.

\begin{table}
\begin{center}
\begin{tabular}{l|cccccc|cccc}
 & $\beta_1$ & $\beta_2$ & $\beta_3$ & $\alpha_1$ & $\alpha_2$ & $h$
       & $b$ & $a$ & $c_1$ & $c_2$   \\  \hline
2.5 \% &  -0.412& -0.205& 0.081 & 0.438 & 0.286 & 0.023  &  -0.219 & 0.319 & -0.061 & -0.033 \\
50  \% &  -0.278& -0.072& 0.240 & 0.530 & 0.374&  0.084 &  0.269 & 0.583  & -0.001  & 0.028  \\
97.5 \%& -0.145 & 0.064& 0.395&  0.621 & 0.463& 0.200  & 0.778 & 0.845 &  0.057&  0.090
\end{tabular} 

\medskip

\medskip

\begin{tabular}{l|ccc|c} 
     &  $\gamma_1$ & $\gamma_2$ &  $\gamma_3$ &  $g$ \\ \hline
2.5\% &  -0.399 & -0.277 & 0.662 & -1.966 \\
50\%   & -0.177 & -0.060 & 0.879 & -0.864  \\
97.5\% &  0.027 &  0.140 & 1.104 &  -0.088 
\end{tabular}

\end{center}
\caption{Posterior quantiles of MCR model parameters for the friendship and
delinquency data. }
\label{table:case2coef}
\end{table}

The results indicate evidence of positive autocorrelation 
for both the network and attribute processes
(represented by $\alpha_1$, $\alpha_2$ and $a$), 
positive homophily with respect to the delinquency attribute ($h$), 
but not not evidence of contagion ($c_1$ and $c_2$).  
This lack of evidence for contagion is in accord with the results
of  \cite{Snijders2010}, who used a stochastic actor-based utility
model to analyze these data. 
Additionally, the posterior distributions of $\bs \beta$  and $\bs\gamma$
indicate evidence of homophily with respect to sex, and that 
males had a higher rate of increase in friendship nominations 
over time. The posterior distributions of $b$  and $g$ 
indicated a lower rate of delinquency among females 
at the beginning of the study ($g$) but 
not a further effect of sex on the delinquency process ($b$).

\section{Discussion}
In this paper we developed a multiplicative co-evolution regression (MCR) model for dynamic network and nodal attribute data, which is able to quantify patterns of autoregression, homophily and contagion in social networks. In the simplest case of a Gaussian network outcome and Gaussian attribute data, the model is essentially a vector autoregressive model. For the more typical case that the network or nodal attribute data are binary or ordinal, we developed a Bayesian approach to parameter estimation and inference. 

This Bayesian approach permits straightforward extensions to the basic MCR model. For example, latent nodal attributes can be included to explain network patterns that is not well-explained by the observed attributes. In this case, we can also model the co-evolution of the network and the latent nodal attributes. 

The work of \cite{Snijders2010} and \cite{Hanneke2010} provide methods for modeling
evolution of network based on network statistics including density, stability, reciprocity and transitivity.
While the MCR model does not require such terms, such effects can be estimated by including network statistics in the regression model. For example, to estimate reciprocity, we include $y_{ji,t-1}$ as a predictor for $y_{ij,t}$.

\bibliographystyle{plainnat}
\bibliography{coevo}

\newpage
\appendix
\section{Gibbs sampler for probit MCR}\label{appendixA}
When $y_{ij,t}$'s are ordinal-valued, we model the network with a probit link and assume $y_{ij,t}=f(z_{ij,t})$. The Gibbs sampler for the probit MCR then include the Steps 1-4 as described in Section 3.2 and also an addition step to update the $z_{ij,t}$'s.

Denote
\begin{equation*}
\left\{
\begin{split}
\vtr{h}_{i,t+1}^{-j}&\equiv \vtr{x}_{i,t+1}-\mg{\theta}_{i}-\mtr{A}^T\vtr{x}_{i,t}-\mtr{C}\mtr{X}_{-j\cdot,t}^T{\vtr{z}_{i,-j,t}},\\
\vtr{h}_{j,t+1}^{-i}&\equiv \vtr{x}_{j,t+1}-\mg{\theta}_{j}^X-\mtr{A}^T\vtr{x}_{j,t}-\mtr{C}\mtr{X}_{-i\cdot,t}^T{\vtr{z}_{-i, j,t}},\\
\mtr{R}_t&\equiv\mtr{M}_t+\mtr{X}_t\mtr{H}\mtr{X}_t^T,\\
\end{split}\right.
\end{equation*}
where $\mtr{X}_{-i\cdot,t}$ refers to the sub-matrix of $\mtr{X}_t$ with the $i$-th row removed and $\mtr{M}_t$ refers to the matrix of $\{\mu_{ij}\}$. Then the equations related to $z_{ij,t}$ include
\begin{equation}
\begin{split}
z_{ij,t}&=\alpha z_{ij,t-1}+r_{ij,t-1}+\epsilon_{ij,t},\\
z_{ij,t+1}&=\alpha z_{ij,t}+r_{ij,t}+\epsilon_{ij,t+1},\\
\vtr{h}_{i,t+1}^{-j}&=z_{ij,t}\mtr{C}\vtr{x}_{i,t}+\vtr{e}_{i,t+1},\\
\vtr{h}_{j,t+1}^{-i}&=z_{ij,t}\mtr{C}\vtr{x}_{j,t}+\vtr{e}_{j,t+1}.\\
\end{split}
\label{equation:latentmodelZijt}
\end{equation}
Given the prior $z_{ij,t}\sim \rm{N}(\mu_0,\sigma_0^2)$, the full conditional distribution of $z_{ij,t}$ is given by $\rm{N}(\mu_{ij,t},\sigma_{ij,t}^2)$, where
\begin{equation*}
\begin{split}
\sigma_{ij,t}^2&= \left(\frac{1+\alpha^2}{\sigma^2}+\vtr{x}^T_{i,t}\mtr{C}^T\mg{\Sigma}^{-1}\mtr{C}\vtr{x}_{i,t}+\vtr{x}^T_{j,t}\mtr{C}^T\mg{\Sigma}^{-1}\mtr{C}\vtr{x}_{j,t}+\frac{1}{\sigma_0^2}\right)^{-1},\\
\mu_{ij,t}&=\sigma_{ij,t}^2\left(\frac{\alpha z_{ij,t-1}+r_{ij,t-1}+\alpha (z_{ij,t+1}-r_{ij,t})}{\sigma^2}+\vtr{x}^T_{i,t}\mtr{C}^T\mg{\Sigma}^{-1}\vtr{h}_{i,t+1}^{-j}+\vtr{x}^T_{j,t}\mtr{C}^T\mg{\Sigma}^{-1}\vtr{h}_{j,t+1}^{-i}+\frac{\mu_0}{\sigma_0^2}\right).\\
\end{split}
\end{equation*}

However, we cannot directly sample from the full conditional distribution due to the restriction of $y_{ij,t}$. Luckily, we only need to restrict our sampling to the interval $[z_{ij,t}^-, z_{ij,t}^+]$, rather than change the full conditionals. The idea for getting the intervals is that the upper bound cannot exceed the minimum value among all the entries of $\mtr{Z}$ whose corresponding entries in $\mtr{Y}$ is higher than $y_{ij,t}$. Similarly, the lower bound is determined by the maximum of those with values in $\mtr{Y}$ lower than $y_{ij,t}$, i.e.,
\begin{equation*}
\left\{
\begin{split}
z_{ij,t}^+&= {\rm min}\{z_{kl,s}: y_{kl,s}> y_{ij,t}; k,l\in\{1,\cdots,m\}, k\neq l, s\in \{1,\cdots,n\}, (k,l,s)\neq (i,j,t)\},\\
z_{ij,t}^-&= {\rm max}\{y_{kl,s}: y_{kl,s}< y_{ij,t}; k,l\in\{1,\cdots,m\}, k\neq l, s\in \{1,\cdots,n\}, (k,l,s)\neq (i,j,t)\}.\\
\end{split}\right.
\end{equation*}
This method is known as rank likelihood, which is introduced in \cite{Hoff2009book}. In this book, the author also presents an alternative way that works for ordinal data with ranks $\{1,\cdots,q\}$. Using this method, we need to update the thresholds $\{h_0,\cdots,h_q\}$ during the Gibbs sampler procedure. Assume a prior for all the thresholds and during the iteration, the threshold $h_s$ is sampled according to the interval $[h_s^-, h_s^+]$, where $h_s^+ = {\rm min}\{y_{kl,s}: y_{kl,s}=s+1\}$ and $h_s^- = {\rm max}\{y_{kl,s}: y_{kl,s}=s\}$. 

To estimate the parameters and latent variables in the model with both ordinal networks and ordinal nodal attributes, we can still follow the Gibbs sampler procedure discussed here, along with an extra step to update $w_{i,k,t}$. In each iteration, we can sample a new point for $w_{i,k,t}$ from its full conditional distribution with restrict to interval $[w_{i,k,t}^-, w_{i,k,t}^+]$. The boundaries/thresholds can be obtained using rank likelihood or sampled from their full conditionals.

\end{document}